\documentclass[aps,prl,twocolumn,floatfix,superscriptaddress,10pt]{revtex4}
\usepackage{graphicx}
\usepackage{amsmath}
\usepackage{amsfonts}
\usepackage{amssymb}
\usepackage{ulem}
\usepackage{color}
\usepackage[T1]{fontenc}
\usepackage[cp1250]{inputenc}
\newcommand{\bea}{\begin{eqnarray}}
\newcommand{\eea}{\end{eqnarray}}
\newcommand{\be}{\begin{equation}}
\newcommand{\ee}{\end{equation}}

\begin{document}

\bibliographystyle{unsrt}


\newcommand{\figwidth}{0.9\columnwidth}
\newcommand{\hc}{\text{H.c.}}

\newcommand{\comment}[1]{\textcolor{red}{#1}}
\newcommand{\commentbis}[1]{\textcolor{blue}{#1}}
\newcommand{\remove}[1]{}
\newcommand{\proposition}[1]{\textcolor{red}{\sout{#1}}}

\title{Rashba induced magnetoconductance oscillations in the LaAlO$_3$-SrTiO$_3$
heterostructure
}
\author{A.~F\^ete}
\author{S.~Gariglio}
\affiliation{D\'epartement de Physique de la Mati\`ere Condens\'ee, University of Geneva, 24 Quai Ernest-Ansermet, 1211 Gen\`eve 4, Switzerland}
\author{A.~D.~Caviglia}
\affiliation{D\'epartement de Physique de la Mati\`ere Condens\'ee, University of Geneva, 24 Quai Ernest-Ansermet, 1211 Gen\`eve 4, Switzerland}
\affiliation{Max-Planck Research Group for Structural Dynamics - Center for Free Electron Laser Science, University of Hamburg, Germany}
\author{J.-M.~Triscone}
\affiliation{D\'epartement de Physique de la Mati\`ere Condens\'ee, University of Geneva, 24 Quai Ernest-Ansermet, 1211 Gen\`eve 4, Switzerland}
\author{M. Gabay}
\affiliation{Laboratoire de Physique des Solides, Bat. 510,
Universit\'e Paris-Sud 11, Centre d'Orsay, 91405 Orsay Cedex,
France}

\begin{abstract}
 We report measurements of the normal state in-plane magnetoconductance in gated
LaAlO$_3$-SrTiO$_3$ samples. As the
orientation of the magnetic field changes  within the plane of the interface,
the signal displays periodic oscillations with respect to the angle between the
field and the direction of the current. We show that in the underdoped to
optimally doped range, a Fermi surface reconstruction takes place due to the
Rashba spin-orbit term and that the oscillations are due to a magnetic field
induced opening and closing of a gap at the $\Gamma$ point for those Ti
out-of-plane orbitals 
having their band minimum close to the Fermi
energy of the system.
\end{abstract}

\maketitle

In response to a strong worldwide demand for powerful portable electronic
devices, engineers have designed novel architectures for processors while
shrinking their sizes and slashing down the required energy input. Current
technology is still mainly silicon-based, but alternative strategies are being
actively pursued which involve materials like graphene or transition metal
oxides \cite{Avouris2010a, Avouris2010b, Zubko2011}. The latter are endowed with
an incredible variety of functional properties which are extremely sensitive to
structural distortions, electronic correlations and crystal chemistry. Recently
a number of novel phenomena at complex oxide interfaces has been discovered 
\cite{Zubko2011} and
among the driving forces that appear to be at play for these new observed
effects are charge confinement/deconfinement in multiple valence
heterostructures. Moreover these artificial structures can help us control 
the carrier flow by  manipulation of the spin.

One promising candidate is the LaAlO$_3$-SrTiO$_3$ (LAO/STO)
heterostructure. Both compounds are band insulators, yet, when a LAO film ($4$ or more unit cells thick) is
deposited layer by layer on top of a (001) STO substrate, the heterostructure
becomes metallic with an electron-like character \cite{Ohtomo,Thiel}. Following the growth procedure defined in Ref. \onlinecite{CancellieriEPL},
 one finds that the conducting sheet is confined to a few
nanometers on the STO side of the interface, and that superconductivity sets in below
a critical temperature $T_c$ in the $0.1$~K range \cite{Basletic2008,Reyren2007}. Carrier
transport was successfully modulated by applying a gate voltage $V$ in top-
\cite{Mannhart2011} and back-gate geometries revealing the existence of a (zero
temperature ($T$)) quantum critical point (QCP) for $V\equiv V_c$ separating an
insulating state from a superconducting state \cite{Caviglia2008} (see Figure 1(a)). At low $T$,
in the normal state, charge transport is well described by $2D$ weak
localization (WL) in a broad range of gate voltages that extends beyond $V_c$
(underdoped regime), but it displays anomalous metallic behavior (AM) for large
enough values of $V$ (optimally doped regime). Upon increasing $V$, a steep
rise of a Rashba-type spin-orbit contribution was evidenced beyond $V_c$, which
appears to track the increase in $T_c$ and the onset of AM above the dome.

In this letter we provide further experimental evidence of the role of the
Rashba spin-orbit coupling on the magnetotransport properties of the LAO/STO
interface. We present transport measurements for
$V>V_c$, and we bring out in full view the remarkable oscillations of the
magnetoconductance (MC) that take place when a magnetic field $B$ is applied
parallel to the interface and then rotated within that plane. We correlate the
onset of the AM with the emergence of a spin-orbit driven Fermi surface reconstruction. In this regime,
as $Ti$ $d_{xz}, d_{yz}$ orbital sub-bands are being filled upon increasing $V$, we
contend that the observed oscillations of the MC are due to the periodic
opening and closing of gaps at the $\Gamma$ point of $Ti$ $d_{xz}$ orbital
sub-bands having their band minimum very close to the Fermi energy. The
amplitude of the oscillations allows one to extract a spin-orbit energy 
in agreement with the WL analysis.

LAO/STO interfaces were prepared by pulsed laser deposition
\cite{Supplementary}. A photolithographic process was used to define Hall bars
for four points DC transport measurements performed at
1.5~K in a cryostat equipped with a rotating
sample probe and a 7~T superconducting magnet. 
Field effect
devices were created using the STO substrate as gate dielectric depositing a
gold contact on the backside. 
The current was applied along the [100] direction and for perpendicular
magnetoconductance the magnetic field was set out of plane along the [001]
direction. For angular magnetoconductance, the magnetic field was applied in
plane and rotated from parallel ($B\parallel$[100], $\phi$=0) to perpendicular
($B\parallel$[010], $\phi$=$\pi$/2) to the current (see Figure S1 in supplementary material).

\begin{figure}[ht]
  \centering
  \includegraphics[width=0.45\textwidth]{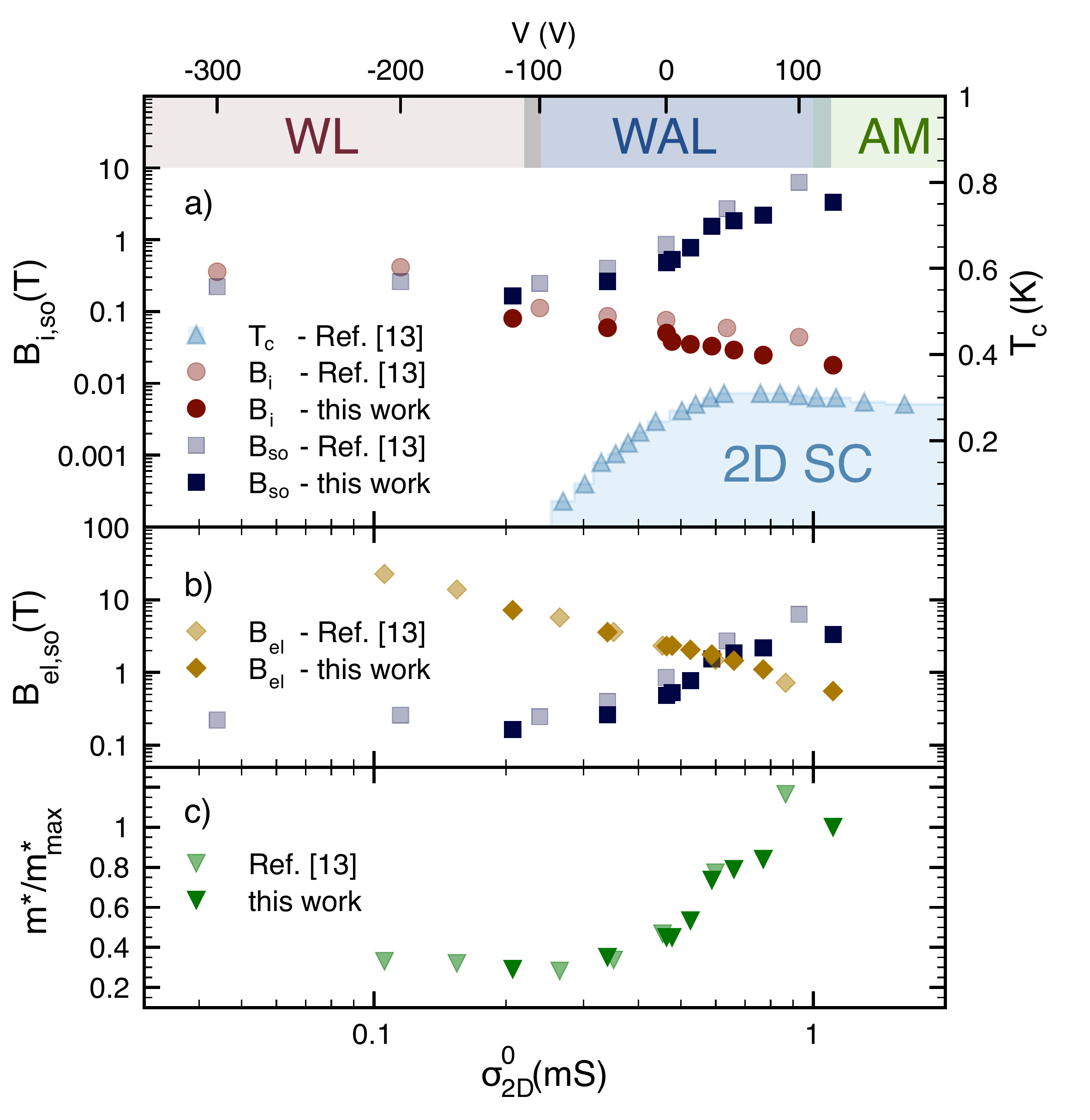}
 \caption{(color online) (a) Evolution of the inelastic ($B_i$) and spin-orbit ($B_{so}$) fields as a function of $\sigma^0_{2D}$ (the sheet conductance at B=0) for the sample of this work and a reference sample
(data from Ref. \onlinecite{Caviglia2010a}) in the diffusive regime. The top voltage scale indicates the corresponding applied voltages for the reference sample. The superconducting phase (2D SC, blue region)
and the transition temperature $T_c$ (triangles) are also shown.
 (b)  Evolution of the elastic  ($B_{el}$) and spin-orbit fields with $\sigma^0_{2D}$ 
highlighting the crossing at $\sigma^0_{2D} \sim 0.6$~mS.  Data points extend slightly beyond the diffusive regime. (c) variation of the transport 
effective mass $m^*$ with $\sigma^0_{2D}$. Here ${m^*}_{max} \sim 2.2 m_e$. WL, WAL, AM shaded regions denote the
weak localization, weak antilocalization and anomalous metal regimes respectively.}
  \label{BiBsoBeM}
\end{figure}

When the applied magnetic field $B$ is perpendicular to the interface, the $B$
dependence of the measured MC is quantitatively captured
using a WL expression for non interacting fermions \cite{Maekawa1981} in a broad
range of gate voltages extending beyond $V_c$. The accuracy of the fit allows us to
estimate the values of the inelastic field, $B_{i}$ and of the spin-orbit field
$B_{so}$ as a function of the gate voltage.

Figure 1 (a) displays the evolution of these two quantities
versus $\sigma^0_{2D}$, 
the sheet conductance (well suited for comparing data from 
different samples).

Previous studies \cite{Caviglia2010a,BenShalom2010a} have shown that the origin
of the spin-orbit part stems from a Rashba interface term. Accordingly,
we can determine the effective Drude mass $m^*$ using the D'yakonov-Perel' relation between the spin-orbit ($\tau_{so}$) and elastic ($\tau$) scattering times $\frac{2\pi}{\tau_{so}}={\Omega_{so}}^2\tau$,
where $\Delta_{so}\equiv \hbar \Omega_{so}=2\lambda E k_F$ is the Rashba energy
($\lambda$ is the material specific Rashba spin-orbit coefficient, $E$ is the interfacial electric field along $z$, the direction normal to the interface,  $k_F$ is the Fermi momentum)\cite{Winkler2003}

\bea \label{effmass}
m^*= \frac{h^2}{4\pi\lambda E}\sqrt{\frac{B_{so}}{\Phi_0}}
\eea
($\Phi_0 =\frac{h}{2e}$) and hence $\tau$ (or $B_{el}$ the elastic magnetic field) 
in the WL regime 
(see supplementary material).

When Eq.(\ref{effmass}) holds, i.e. in the range of $V$ 
such that spin-orbit terms contribute to diffusive processes, we consider that $\lambda$ has little $V$ 
dependence. By contrast, changing the
gate voltage modifies the sheet carrier density $n_S$ and causes a variation of $E$. The change in $E$
 can be determined from an electrostatic model of the confined electron gas \cite{note1}, 
taking into account the electric field dependence of the STO dielectric constant (see Ref. \onlinecite{Ueno2008}). 

Using these prescriptions, we present the variations of $B_{el}$,  $B_{so}$ and $m^*$ with
$\sigma^0_{2D}$ (or V) in Figures 1 (b) and (c). To highlight the trends,
 we include data points slightly beyond the diffusive regime (see below). We note that $m^*$ is $\sim
3.5 $ lighter in the underdoped regime than in the optimally doped regime, which
is consistent with the picture of an interface charge transport evolving from being
 $d_{xy}$-dominated ($m^*\sim 0.64 m_e$ \cite{Santander2011}, $m_e$ is the bare electron mass)
to being $d_{xz}/d_{yz}$-dominated ($m^*\sim 2.2 m_e$),
upon increasing $V$ beyond $\sigma^0_{2D}\sim0.5$~mS \cite{Seo2009,
BenShalom2010a, Caviglia2010a,BenShalom2010b,Caviglia2010b, Copie2009,Pentcheva2009}.  
The concomittant rise in $\tau$ (Figure S2
 supplementary material) fits with this picture, since it signals that additional conduction channels start contributing.
As reported in
Refs. (\cite{Caviglia2010a,BenShalom2010a}), $\tau_{so}$ 
decreases steeply across $V_c$, to the extent that $\tau_{so} \sim \tau$ for $\sigma^0_{2D}
\sim 0.6$~mS (Figure \ref{BiBsoBeM} (b)).  When this occurs 
spin-orbit
processes cease to be diffusive. The spin-orbit time is no longer given by the
D'yakonov-Perel' expression and the Rashba term becomes a \textit{bona fide} new energy
scale in the problem, on equal footing with the kinetic part. Band structure
needs to be recalculated in the presence of the (Fermi surface reconstructing)
Rashba hamiltonian. The evolution towards this AM regime is suggestive of a scenario of spin-orbit protected transport 
(against disorder) 
in the 2D conducting sheet.

\textit{Ab-initio} density functional theory (DFT) band structure
calculations have been carried out for the LAO/STO system \cite{Popovic2008,
Pentcheva2009, Son2009, Ghosez2011}. These studies reveal that the conduction
band consists of a manifold of $t_{2g}$ states originating from  a decomposition
of the $t_{2g}$ triplet into sub-bands, in response to the interfacial confining
electric field \cite{Ueno2008, Shen2011, Santander2011}. The energy minima of
these sub-bands are located at the $\Gamma$ point; the lower the value of the
energy at the minimum the closer to the interface along $z$ the wavefunction is. According to Ref. (\cite{Ghosez2011}), for values of $n_S$
$\sim 10^{14}$~cm$^{-2}$ the states which lie closest to the interface, i.e. which do not extend beyond
the first 3 unit cells from the STO boundary, on the STO side, have $d_{xy}$
symmetry and account for a large part of the carrier concentration. $d_{xz}$, $d_{yz}$ sub-bands spread out deeper
into the (STO) bulk, and, as they are being filled, they contribute to $n_S$ on the order of 
$10^{13}$~cm$^{-2}$. 
These out of plane orbitals lie close the Fermi surface and we argue below that
they control charge transport in the AM regime.


\begin{figure*}[ht]
\begin{center}$
\begin{array}{cc}
\includegraphics[width=0.45\textwidth]{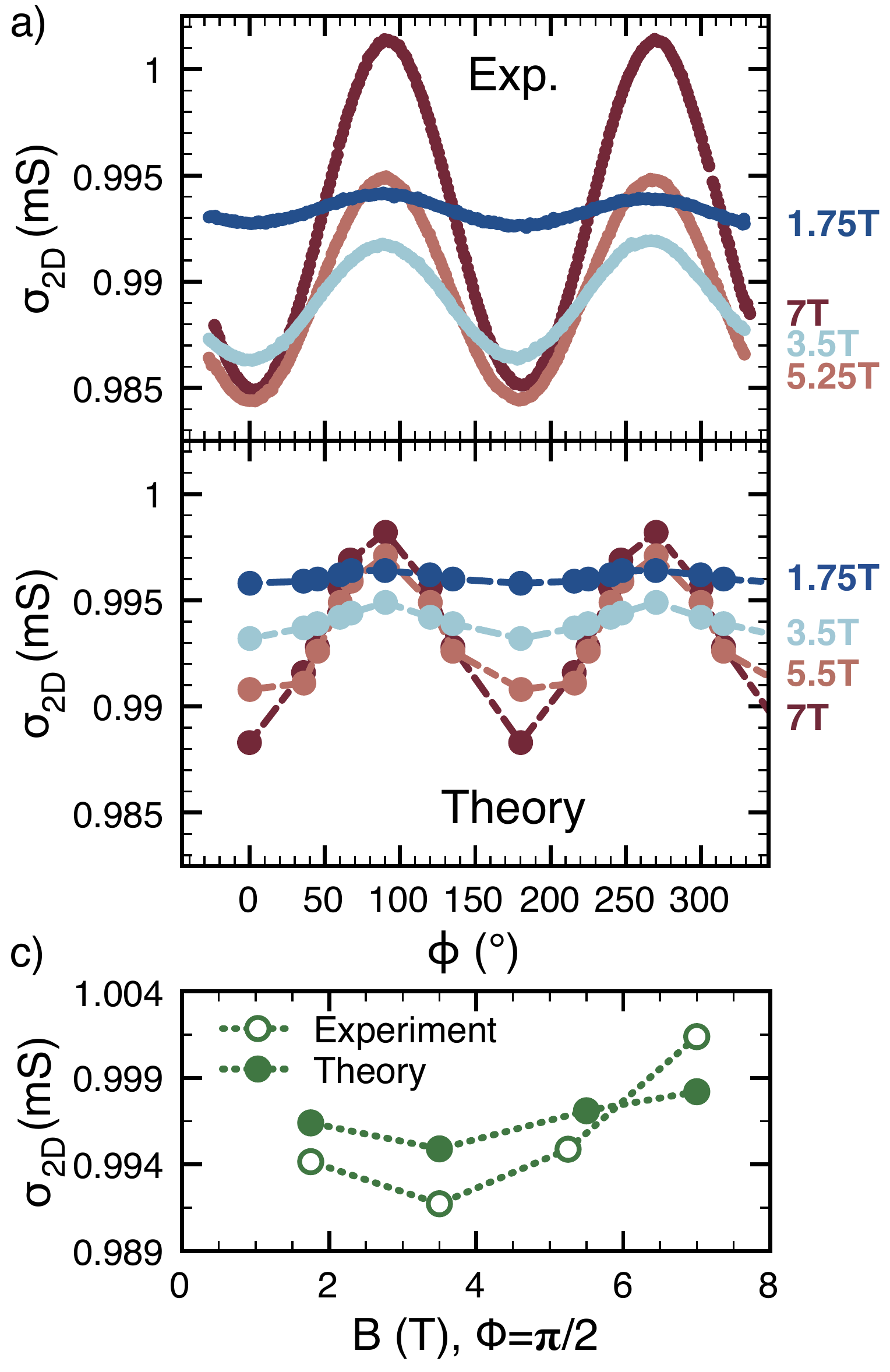} &
\includegraphics[width=0.45\textwidth]{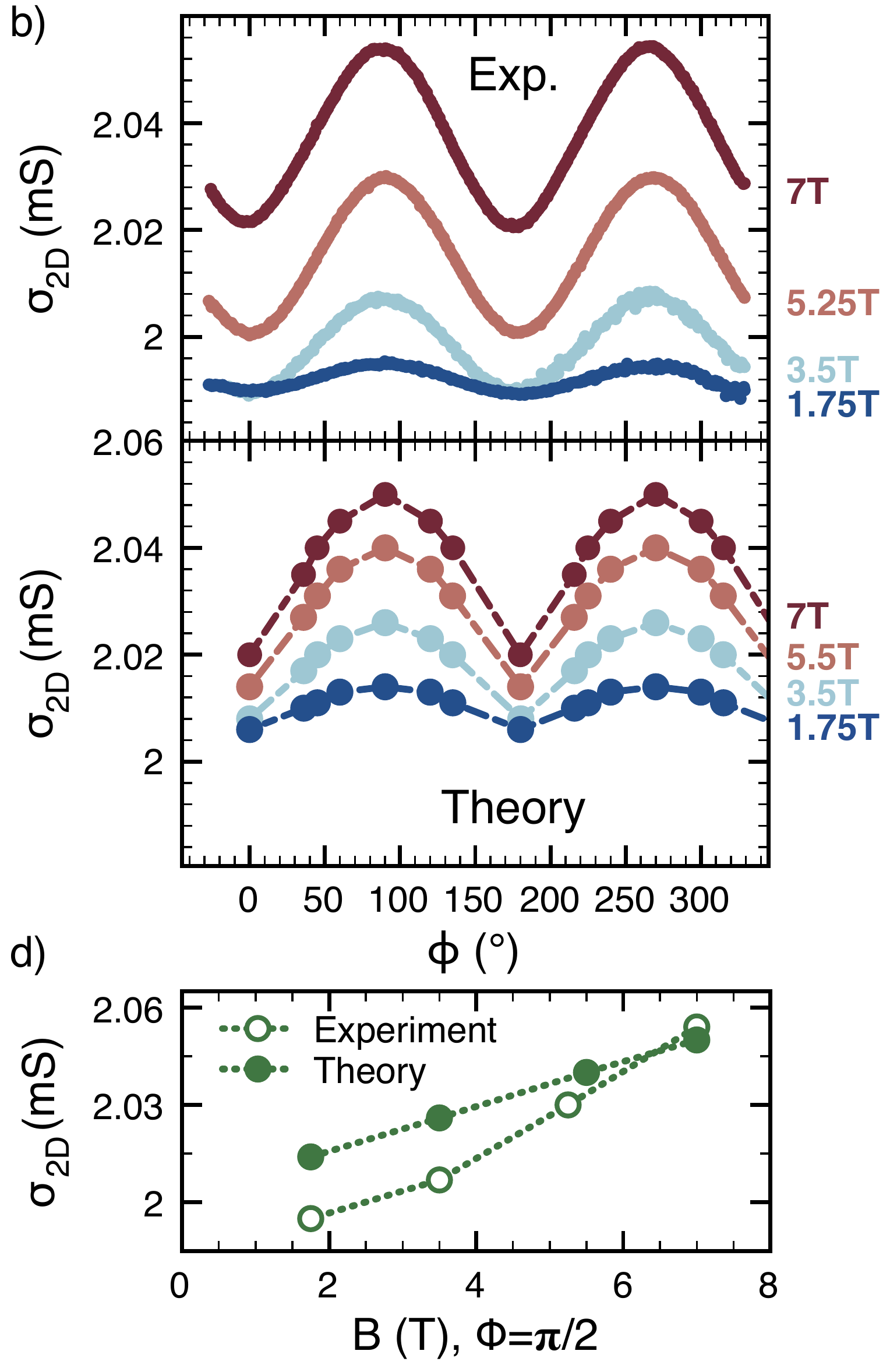}
\end{array}$
\end{center}
\caption{(color online) (a) and (b) Experimental and theoretical plots of the conductance $\sigma_{2D}(B,\phi)$ versus
$\phi$, the  angle between the in-plane magnetic field $B$ and the current for
various values of $B$. (a) corresponds to $\sigma^0_{2D}=1$~mS and (b) to $\sigma^0_{2D}=2$~mS. (c) and (d) are experimental and theoretical dependence of the conductance on the in-plane magnetic field for $\phi$=$\pi /2$.
\label{DeltasigmaVSphi}
}
\end{figure*}

To showcase the significant role played by the $d_{xz}$, $d_{yz}$ states, we measure the angular MC of the 2DEG
when a magnetic field is applied in plane and rotated from parallel 
to perpendicular to the current. In this geometry the magnetotransport is not affected by orbital contributions.
Figure \ref{DeltasigmaVSphi} shows the variation of the conductance with
the angle $\phi$ for four values of the magnetic field when $\sigma^0_{2D}$ is 1~mS
(a)) and $\sigma^0_{2D}$ is 2~mS (b)). Maxima of the conductance are seen
for both dopings when the external magnetic field is applied perpendicularly to
the current. However the maximum of the oscillation $\sigma_{2D}(B,\phi=\pi/2)$ evolves in a non monotonic way for the first doping which
is not case for the higher conductance state. Note that oscillations of the MC in the parallel field geometry have been reported by Ben Shalom et al (\cite{Dagan2009}).
For the range of sheet conductances and the field intensity that they considered (typically higher than ours) 
they found a positive MC for $\phi=0$, and suggested that a magnetic order forms at the interface. 

For the doping range considered here, we may understand the behavior of the parallel MC as the field is rotated within
the {\bf xy} plane if we note that carriers in the $d_{xz}$ or $d_{yz}$
sub-bands have one light and one heavy mass (${m}_h\sim 20 m_e$)
\cite{Santander2011, Seo2009}. Thus, qualitatively, the  $d_{xz}$ and $d_{yz}$
orbitals are 1D-like and, since current flows along {\bf x}, we focus on the
former type which gives a higher contribution to transport. In the presence of
the Rashba term, the spin-split bands exhibit an energy gap at the $\Gamma$
point when $B$ is along {\bf x} ($\phi=0$) and a Zeeman-like offset when
$B$ is along {\bf y} ($\phi=\frac{\pi}{2}$) (Figure S3
 supplementary material). The impact of this effect on
transport depends on the position of $E_F$ relative to $E_{\Gamma}$, the energy
of the electronic states at the $\Gamma$ point when $B = 0$, since the density of states (DOS) at the Fermi energy,
$g(E_F)$, enters the expression of $\sigma_{2D}$. The conductivity will thus show a dip 
for $\phi=0$, provided $E_F\sim E_{\Gamma}$~. If $E_F$ is not close to $E_{\Gamma}$,
$g(E_F)$ is almost unchanged as compared to its $B = 0$ value. As mentioned
in the previous section, in the underdoped to optimally doped range, $E_{\Gamma}$ of most of the
sub-bands with a $d_{xy}$ character is well below $E_F$. Yet, the $E_{\Gamma}$ of  $d_{xz}$ symmetry bands
 are close to $E_F$. So the Rashba induced modulation of $g(E_F)$ is controlled by the 
change in the $d_{xz}$ carrier DOS. Insofar as
one may reasonably assume that $\tau$ does not depend on $B$ in the 2D case,
one expects a periodic variation of the MC with $\phi$, displaying crests for
$B$ along {\bf y} and troughs for $B$ along {\bf x}.

\begin{figure}
      \includegraphics[width=0.45\textwidth]{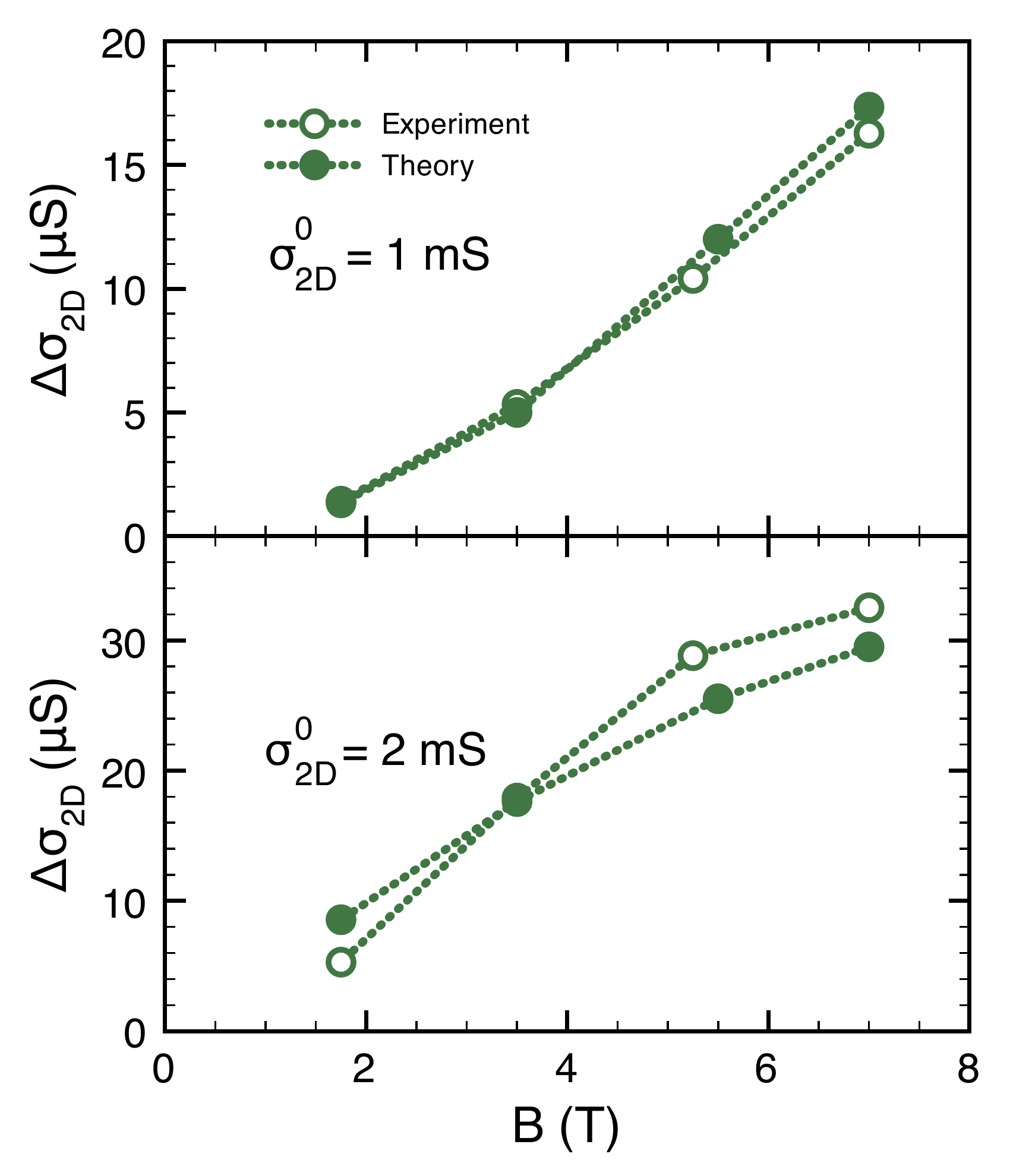}
  \caption{(color online) Experimental and model determined plots of $\Delta\sigma_{2D}$ versus $B$ for $\sigma^0_{2D}=1$~mS and $\sigma^0_{2D}=2$~mS.}
  \label{DeltasigmavsB}
\end{figure}

In the framework of this 1D picture and with the additional simplifying
assumption that the variation of $\sigma_{2D}$ with $\phi$ is entirely due to the
$d_{xz}$ sub-band closest to $E_F$, we find 
$(\sigma_{2D}(B,\phi=\pi/2)-\sigma_{2D}(B,\phi=0))/\sigma_{2D}(B,\phi=0)=1/8
(\Delta_{so}/E_F)^2$
when the Zeeman energy is larger than the condensation energy of the Rashba
state ($\Delta_{so}$) (see also \cite{Castellani2001}).
Beyond the qualitative 1D model, we have modelled the evolution of the band
structure for the $t_{2g}$ orbitals in the applied $B$. We use a tight binding model 
featuring kinetic and Rashba terms and we take into account
the finite value of ${m}_h$ and the anisotropy of $\lambda$ in the {\bf xy}
plane for the $d_{xz}, d_{yz}$ orbitals.  The mobility, $\lambda$ and the
gyromagnetic factor $g$ all depend on $V$, but we consider that they do not
change appreciably with $\phi$ nor with the magnitude of $B$ in our experiments.
For a given $V$, the variation of $\delta n$ -- i.e. the change  in the 
$d_{xz}, d_{yz}$ carrier concentrations -- with $\phi$ and $B$ depends on the
values of the spin-orbit and of the Zeeman energies. The
conductance $\sigma_{2D}(B,\phi)$ is then proportional to $\delta n$. 
Experimental data and plots obtained from the model are shown 
in Figures \ref{DeltasigmaVSphi} (c) and (d). We note
that while $\sigma_{2D}(B,\phi)$ increases monotonically with $B$ at fixed $\phi$ for
$\sigma^0_{2D}=2$ mS, such is not the case for $\sigma^0_{2D}=1$ mS, a feature
which is correctly captured by our model. Figure \ref{DeltasigmavsB} 
displays the evolution of 
$\Delta\sigma_{2D}=\sigma_{2D}(B,\phi=\pi/2)-\sigma_{2D}(B,\phi=0)$ versus $B$ 
according to our model and the experimental results. We find good agreement using $\Delta_{SO}= 7 (2.5)$~meV 
for $\sigma^0_{2D}= 2 (1)$~mS respectively. These
values fall within the range of previous experimental estimates \cite{Caviglia2010a,
BenShalom2010a}. 

\begin{figure}
      \includegraphics[width=0.45\textwidth]{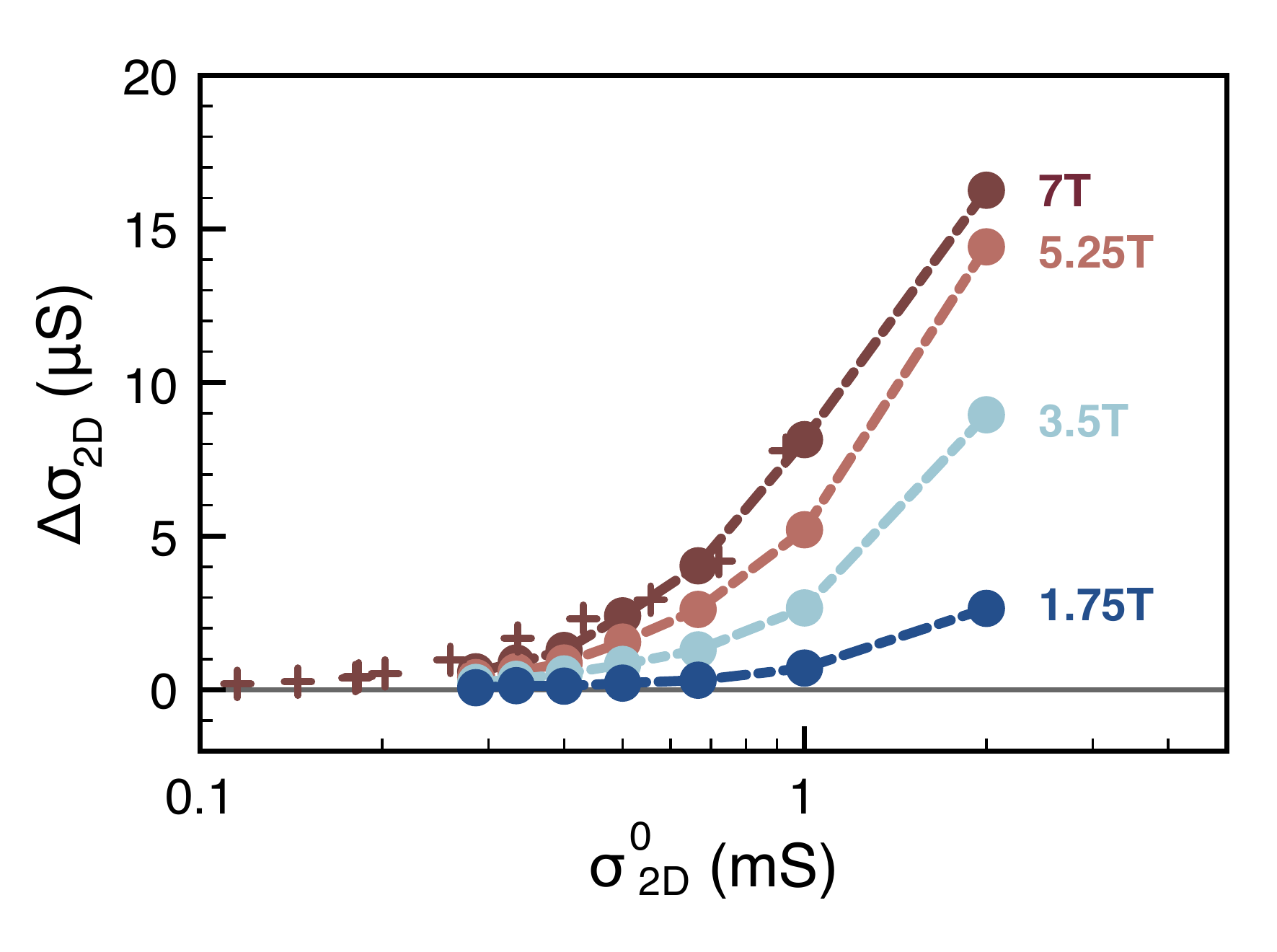}
  \caption{(color online) Evolution of the amplitude of the experimental oscillations $\Delta\sigma_{2D}$ ($\Delta\sigma_{2D}=\sigma_{2D}(B,\phi=\frac{\pi}{2})-\sigma_{2D}(B,\phi=0)$)
 as a function of $\sigma^0_{2D}$ . "Plus" symbols pertain to measurements performed on an additional sample at 7~T.}
  \label{Deltasigmavssigma2D}
\end{figure}

Figure \ref{Deltasigmavssigma2D} shows the evolution of the experimental oscillation amplitude 
$\Delta\sigma_{2D}$
as  a function of $\sigma^0_{2D}$ for different magnetic fields. As can be seen, $\Delta\sigma_{2D}$ tends to zero for a sheet 
conductance in the $0.1 - 0.3$~mS range. These sheet conductance values -- which lie in the diffusive regime -- are close to 
the QCP, suggesting a potentially important role played by the $d_{xz}, d_{yz}$ orbitals in establishing superconductivity.

In summary, our findings underscore the evolution that takes place in the LaAlO$_3$-SrTiO$_3$ heterostructure, as one tunes the
gate voltage in the range where superconductivity is observed at low temperature. For low $V$, conduction
is diffusive and is dominated by the $d_{xy}$ orbitals, as the impact of disorder is expected to be more severe for the 1D-like $d_{xz}, d_{yz}$ states. For larger values of
$V$, the nature of transport changes and the out of plane $d_{xz}, d_{yz}$ orbitals start contributing to the zero field conductivity,
as evidenced by the evolution of $m^*$, $B_{el}$, $B_{so}$ and by the oscillations of the in-plane magnetoconductance. 
For these orbitals, the effect of the strong spin-orbit interaction has to be taken into account 
at the band structure level for the calculation of their contribution to the transport.

We would like to thank M. Lopes and S. C. M\"uller for their technical assistance. This work was supported by the Swiss National
Science Foundation through the National Center of Competence in Research,
Materials with Novel Electronic Properties, MaNEP and division II, by the Institut
Universitaire de France (MG), and  the
European Union through the project OxIDes.

\onecolumngrid
\newpage
\part{Supplementary Information\newline}

\setcounter{figure}{0}

\section{Sample preparation and experimental geometry}\label{section1}

A LaAlO$_3$ (LAO) layer thicker than 4 unit cells was grown on a (001)
TiO$_2$-terminated SrTiO$_3$ (STO) substrate heated at 800$^\circ$C in an oxygen
pressure of 10$^{-4}$ mbar. The KrF excimer laser fluency was set to 0.6
J/cm$^2$ with a repetition rate of 1 Hz. After deposition, the oxygen pressure
was raised to 0.2 bar and the sample was kept at a temperature of about
530$^\circ$C during 1 hour before cooling to room temperature. The layer
thickness was estimated \textit{in situ} from intensity oscillations of the
reflection high energy electron diffraction (RHEED) spots.

\begin{figure}[ht]
 \centering
  \includegraphics[width=0.95\textwidth]{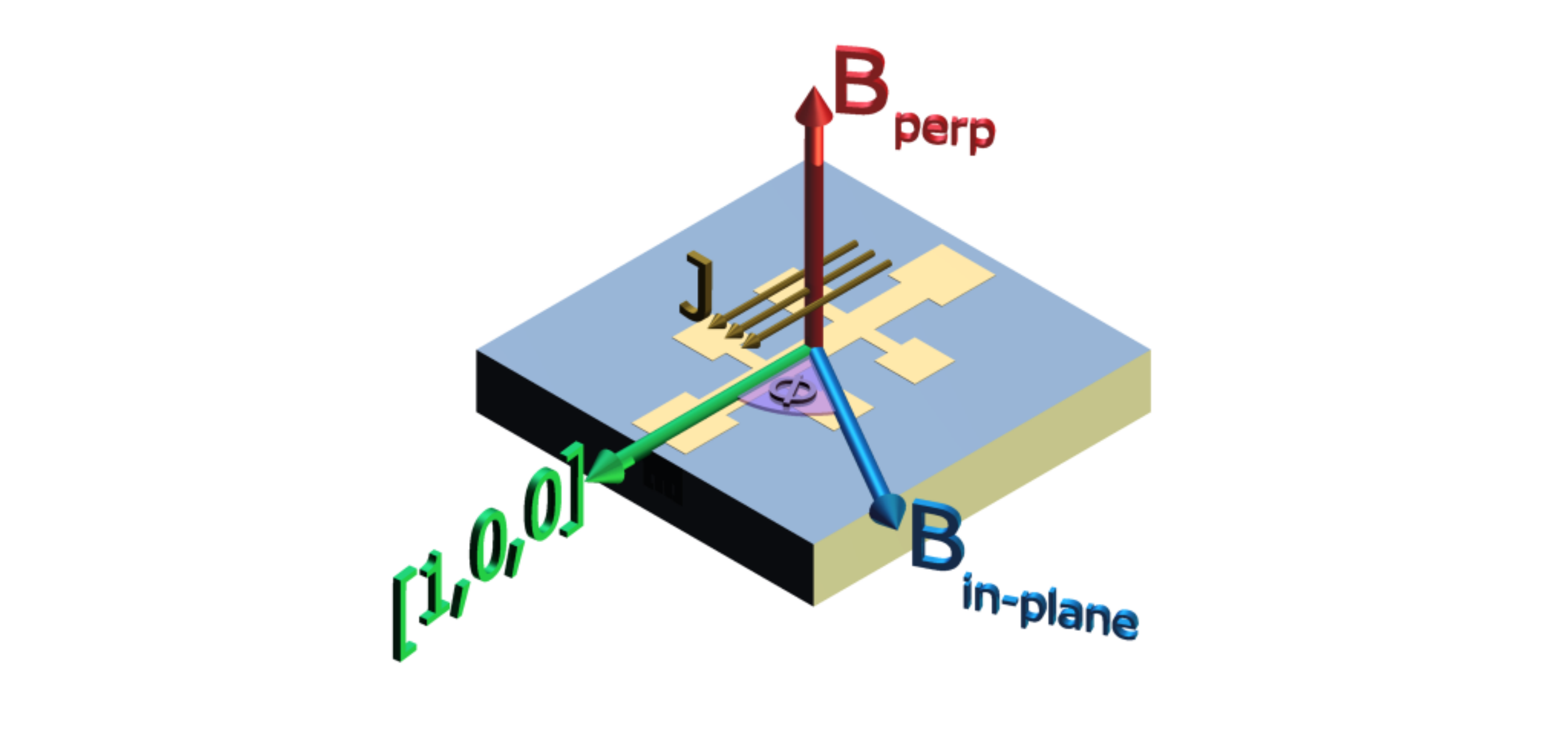}
  \caption{Shematic view of a field effect device. 
$\phi$ is the angle between the current ($J$) and the in-plane magnetic field $B_{in-plane}$.
}
 \label{Geometries}
\end{figure}

\section{Weak localization analysis}

In the diffusive regime, the perpendicular magnetoconducance (MC) is quantitatively 
captured using weak localization corrections. 
We have estimated the spin-orbit field $B_{so}$  and the inelastic field $B_{i}$
as a function of the gate voltage using the Maekawa-Fukuyama expression for
the magnetoconductance \cite{Maekawa1981}. In order to obtain both the effective mass
$m^*$ and the bare (Drude) elastic transport time $\tau$, we need to determine the bare conductance ($\sigma^0_D$) for various $V$. To that end,
we use the experimentally measured $B=0$ sheet conductance ($\sigma^0_{2D}$) and
substract the zero field WL correction to the conductance, given by 
\bea\label{HLN}
\Delta\sigma^0_{WL}=\frac{e^2}{\pi h}\left ( Log[(1+\frac{B_{so}}{B_{i}})(1+\frac{2
B_{so}}{B_{i}})^{\frac{1}{2}}]-Log[\frac{B_{el}}{B_{i}}] \right )
\eea
\subsection{Effective mass}\label{section2}
In order to determine $m^*$, we
consider the D'yakonov-Perel' relation \cite{Dyakonov1972} between
$\tau_{so}$ and $\tau$, namely
$\frac{2\pi}{\tau_{so}}=\Omega_{so}^2\tau$ where $\Delta_{so}\equiv \hbar
\Omega_{so}=2\lambda E k_F$ is the Rashba energy
($E$ is the interfacial electric field, $k_F$ is the Fermi momentum).

In the diffusive regime, the spin-orbit field is defined as $B_{so}=\frac{\Phi_0}{4\pi
D\tau_{so}}$, where $\Phi_0=\frac{h}{2e}$ and $D=\frac{1}{2}{v_F}^2\tau$ is the Drude (bare) diffusion constant.
Using the D'yakonov-Perel' relation, it follows that $D\tau_{so}=\frac{1}{2}{v_F}^2\frac{2\pi}{\Omega_{so}^2}$. 
Taking into account the fact that $m^*v_F=\hbar k_F$ and that  $\Omega_{so}=\frac{2\lambda E k_F}{\hbar}$ we obtain
\bea
\frac{\Phi_0}{B_{so}}=4\pi D\tau_{so}=\frac{h^4}{4\pi^2}\frac{1}{(\lambda E m^*)^2} \nonumber
\eea
 Hence
\bea \label{effmass}
m^*= \frac{h^2}{4\pi\lambda E}\sqrt{\frac{B_{so}}{\Phi_0}}
\eea
\subsection{Elastic scattering time}\label{section4}
\begin{figure}[ht] 
 \centering
  \includegraphics[width=0.45\textwidth]{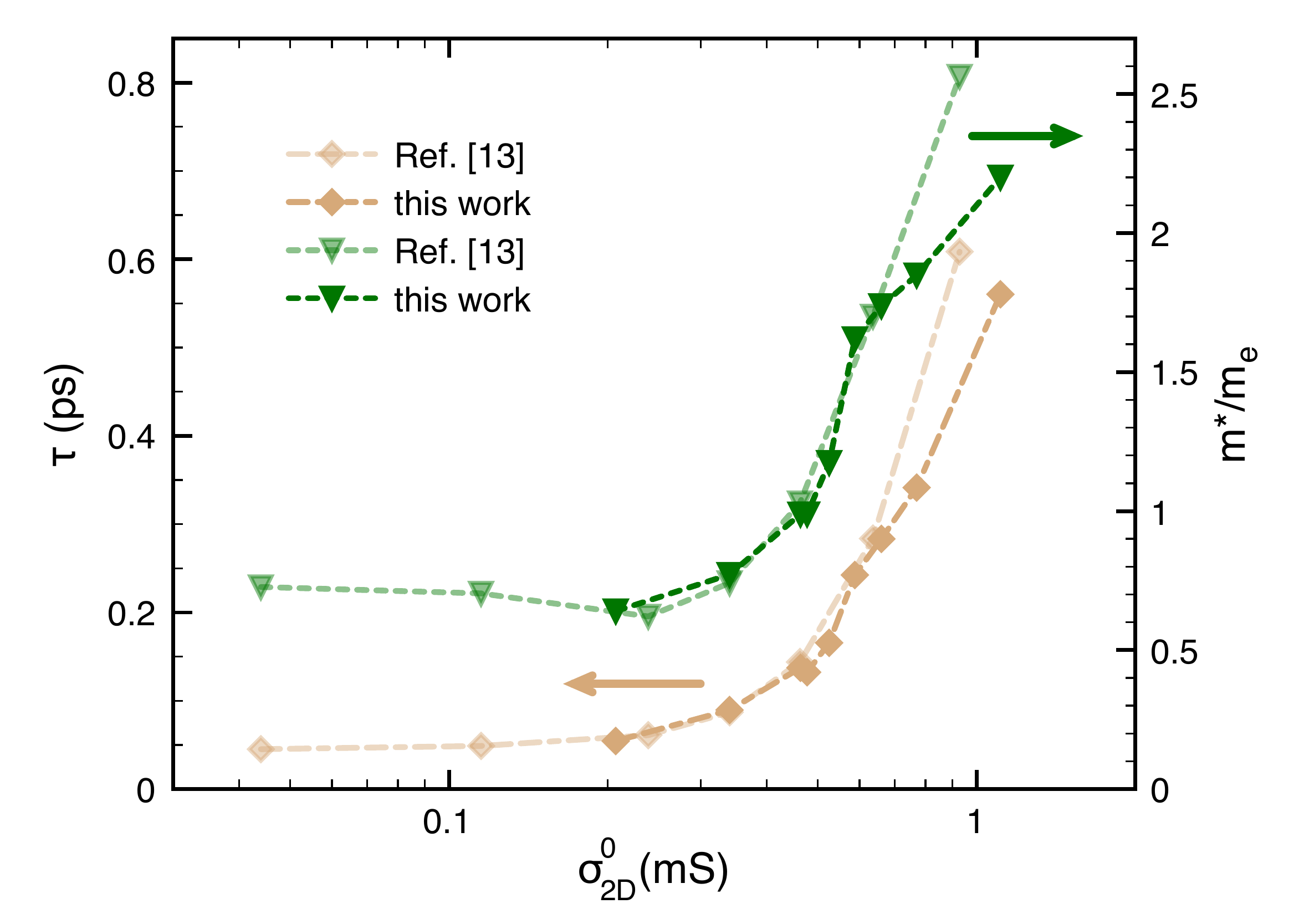}
  \caption{Dependence of the elastic scattering time (left scale) and of the effective mass
 (right scale) on the sheet conductance $\sigma^0_{2D}$.}
 \label{tauevssigma}
\end{figure}

The elastic field in Eq.(\ref{HLN}) is given by $B_{el}=\frac{\Phi_0}{4\pi D\tau}$. Using 
$D=\frac{1}{2}{v_F}^2\tau$, $m^*v_F=\hbar k_F$, and $k^2_F=2\pi n$ allows us to rewrite $B_{el}$ in terms of the
Drude mobility $\mu_D=\frac{e\tau}{m^*}$ and of the mobile sheet carrier density $n$.
Because $\Delta\sigma^0_{WL}$ is a correction to the conductivity and because
$B_{el}$ enters Eq.(\ref{HLN}) through a Log term, we may replace $\mu_D$ by the
renormalized $\mu$ -- obtained from measuring the sheet conductivity -- in the
expression of $B_{el}$ 
when we compute $\Delta\sigma^0_{WL}$. 
We then subtract $\Delta\sigma^0_{WL}$
 from the measured sheet conductance $\sigma^0_{2D}$ to get
the Drude conductivity $\sigma^0_{D}=n e \mu_D$. Besides, magnetotransport allows us to extract the sheet carrier density and the effective mass
(Eq.(\ref{effmass})); from these three quantities we determine the elastic scattering time $\tau$ and 
its variation with $\sigma^0_{2D}$ is shown in Figure \ref{tauevssigma}. We note the steep rise in $\tau$
 which mirrors the sharp increase in $m^*$ beyond $\sigma^0_{2D}\sim 0.5$ mS. Such upturn may be understood
if we recall that while $d_{xy}$ orbitals appear to dominate transport for lower values of $\sigma^0_{2D}$, $d_{xz,yz}$ 
states contribute significantly for larger $\sigma^0_{2D}$.The steep rise suggests that the mobility of the
 $d_{xz,yz}$ orbitals is sizably larger than that of the $d_{xy}$ orbitals. 

\section{Band structure model for in-plane field magnetoconductance}\label{section3}

\begin{figure}[ht]
 \centering
  \includegraphics[width=0.45\textwidth]{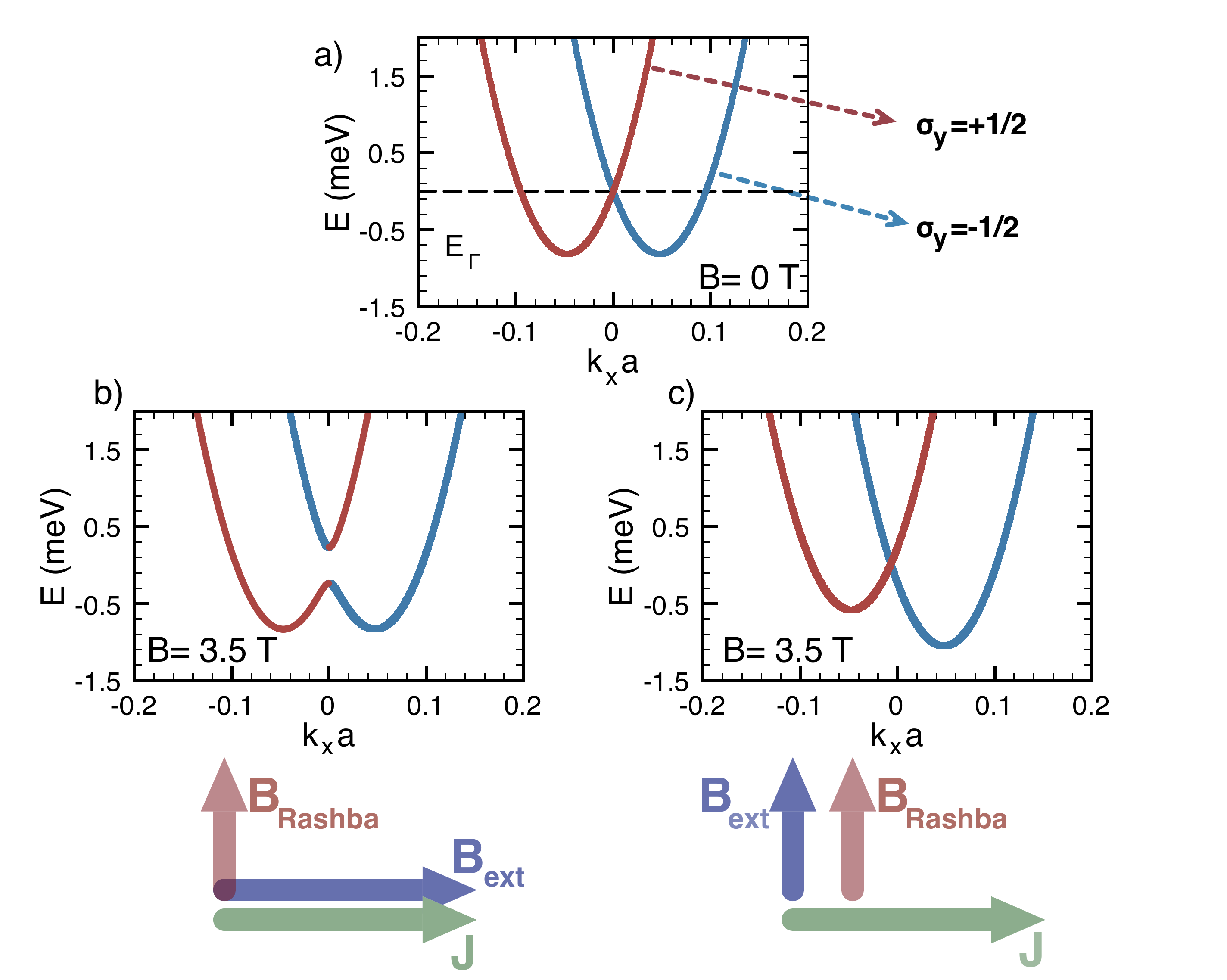}
  \caption{Shematic view of the band structure according to our model. a) We
show a
$d_{xz}$ sub-band split by a spin-orbit coupling of 5~10$^{-12}$~eV~m. The
effective mass used in the calculation is m$_{x,xz}^*$=0.64 m$_e$. $E_{\Gamma}$
defines the crossing point of the spin-split bands. b) Applying
a magnetic field parallel to the current direction ([100] direction, $\phi=0$)
opens a gap at the gamma point: when the Fermi level is at $E_{\Gamma}$, the
change
in carrier density leads to a decrease of conductance. c) When the magnetic
field is in plane and
perpendicular to the current ([010] direction, $\phi=\pi/2$), a Zeeman effect is
observed on the band
structure.}
 \label{BandStrucutre}
\end{figure}

Beyond a characteristic voltage $V> V_c$, $\tau_{so}$ can become less
than $\tau$ -- the transport elastic time
-- implying that the spin-orbit term is no longer diffusive but that it changes
the eigenenergies of the states. Figure \ref{BandStrucutre} displays the band
structure for the $d_{xz}$ orbitals split by the Rashba coupling.

\newpage

\section{Modelling parallel-field magnetoconductance oscillations}
In order to describe the evolution of the 2DEG conduction bands as the orientation of 
the parallel field $B$ changes in the plane of the interface, we use a tight-binding 
representation of the $t_{2g}$ triplet states. In momentum space, 
the Hamiltonian has diagonal entries ${\mathbf H_k}$ consisting of $2\times 2$ spin blocks;
 ${\mathbf H_k}$ contains a kinetic part 
 \bea \label{tightbind}
 \left(2 t_x(1- \cos{(k_x d)}) +2t_y(1- \cos{(k_y d)})\right) \mathbb{1}_2
\eea
 a Rashba piece
\bea
\frac{\alpha_{y R}}{d}\sin{(k_y d)}\;\sigma_x -\frac{\alpha_{x R}}{d}\sin{(k_x d)}\;\sigma_y
\eea

and a Zeeman term
\bea
-\frac{g\mu_B}{2}B\cos{\phi}\;\sigma_x -\frac{g\mu_B}{2}B\sin{\phi}\;\sigma_y
\eea

$\mathbb{1}_2$ is the unit matrix in spin space and $\sigma_x$, $\sigma_y$, $\sigma_z$ are the Pauli matrices. 
The angle between the in-plane magnetic field $B$
 and the $x$ (current) direction is $\phi$. $d$ ($\sim 3.91$\AA) is the lattice spacing.

$d_{xy}$ orbitals are invariant under a $x\leftrightarrow y$ symmetry, so that $t_x =t_y$ and 
$\alpha_{x R}=\alpha_{y R}=\alpha_R$. By contrast, $d_{xz}$ and $d_{yz}$ states are anisotropic and, 
from the small $k$ limit, $t_y /t_x$ ($t_x /t_y$) is equal to the ratio of the light mass $m_l$ 
over the heavy mass $m_h$ for the $d_{xz}$ ($d_{yz}$) orbitals respectively. Similarly, 
the coefficients of the Rashba contributions are anisotropic and, following the discussion presented 
in Ref. [\onlinecite{Simon2010}], we account for this by setting 
$\alpha_{y R}/\alpha_{x R}=m_l/m_h $
 ($\alpha_{x R}/\alpha_{y R}=m_l/m_h $) for the $d_{xz}$ ($d_{yz}$) orbitals
 respectively. 
\bea
E_{\pm}= 2 t_x(1- \cos{(k_x d)}) +2t_y(1- \cos{(k_y d)})\pm\sqrt{(\frac{\alpha_{x R}}{d}\sin{(k_x d)}+\frac{g\mu_B}{2}B\sin{\phi})^2+(\frac{\alpha_{y R}}{d}\sin{(k_y d)}-\frac{g\mu_B}{2}B\cos{\phi})^2}
\eea
 
To the above energy we must add the energy term describing the confinement in the 
direction perpendicular to the interface. It produces an {\it a priori} different value of 
the energy $E_{\Gamma}$ at the $\Gamma$ point for each of the $d$ states \cite{Santander2011}.
With these energies, we may determine the occupations of the orbitals for a given value of $E_F$.
When current flows in the $x$ 
direction, we focus on the $d_{xz}$ orbital whose $E_{\Gamma}$ is close to $E_F$. 
 If the elastic transport time does not change appreciably upon varying $B$ and $\phi$, 
  we claim that
the angular dependence of the magnetoconductance stems from the modulation of the DOS as $B$ 
rotates within the plane of the interface. So we write $\sigma_{2D}(B,\phi)=\sigma^{(1)}(n)+ C(n)\delta n$.
$\sigma^{(1)}(n)$ and $C(n)$ depend on the mobile carrier concentration $n$ (on the gate voltage) 
but not on $B$ or $\phi$.  The carrier density in the $d_{xz}$ band whose $E_{\Gamma}$ is close to $E_F$, 
$\delta n$ ($<<n$), changes with the gate voltage, $B$ and $\phi$. 

$\sigma^{(1)}(n)$ is the contribution to the conductance of those $d$ states whose $E_{\Gamma}$ 
is not close to $E_F$. As explained in the text, we expect negligible $B$ and $\phi$ dependence in that case.

For a given $\sigma^0_{2D}$ we choose one value of the field 
($7$T for $\sigma^0_{2D}=2$mS, $5.5$T for $\sigma^0_{2D}=1$mS) 
and we determine $\sigma^{(1)}(n)$, $C(n)$, $\alpha_R$ and $g$ 
so as to best fit the experimental variation of $\sigma_{2D}(B,\phi)$ with $\phi$.
 We subsequently use these same values when we change $B$ and compute the new $\sigma_{2D}(B,\phi)$.
 A comparison between the theoretical and the experimental curves is shown in Figures 2 and 3.

\bibliographystyle{prsty}

\begin{thebibliography}{10}

\bibitem{Avouris2010a}
Ph.~Avouris, Nano Lett. {\bf10}, 4285 (2010).

\bibitem{Avouris2010b}
F.~Xia, D.~B.~Farmer, Y.-M.~Lin, Ph.~Avouris, Nano Lett. {\bf10}, 715 (2010).

\bibitem{Zubko2011}
P.~Zubko, S.~Gariglio, M.~Gabay, P.~Ghosez, J.-M.~Triscone, Annu. Rev. Condens. Matter Phys. {\bf2}, 141 (2011).

\bibitem{Ohtomo}
A.~Ohtomo, H.~Y.~Hwang, Nature {\bf427}, 423--426 (2004).

\bibitem{Thiel}
S.~Thiel, G.~Hammerl, A.~Schmehl, C.~Schneider, J.~Mannhart, Science {\bf313}, 1942 (2006).

\bibitem{CancellieriEPL}
C.~Cancellieri, N.~Reyren, S.~Gariglio, A.~D.~Caviglia, A.~F\^ete, J.-M.~Triscone, EPL {\bf91}, 17004 (2010).

\bibitem{Basletic2008}
M.~Basletic \textit{et~al.}, Nat. Mat. {\bf7}, 621-625 (2008). 

\bibitem{Reyren2007}
N.~Reyren \textit{et~al.}, Science {\bf317}, 1196-1199 (2007).

\bibitem{Mannhart2011}
L.~Li, C.~Richter, S.~Paetel, T.~Kopp, J.~Mannhart, R.C.~Ashoori, Science {\bf232}, 825 (2011).

\bibitem{Caviglia2008}
A.~D.~Caviglia \textit{et~al.}, Nature {\bf456}, 624-627 (2008).

\bibitem{Supplementary}
See supplementary informations for thin~films growth and characterization.

\bibitem{Maekawa1981}
S.~Maekawa, H.~Fukuyama, J. Phys. Soc. Jpn. {\bf50}, 2516 (1981).

\bibitem{Caviglia2010a}
A.~D.~Caviglia, M.~Gabay, S.~Gariglio, N.~Reyren, C.~Cancellieri, J.-M.~Triscone, Phys. Rev. Lett. {\bf104}, 126803 (2010).

\bibitem{BenShalom2010a}
M.~Ben~Shalom, M.~Sachs, D.~Rakhmilevitch, A.~Palevski, Y.~Dagan, Phys. Rev. Lett. {\bf104}, 126802 (2010).

\bibitem{Winkler2003}
R.~Winkler, \textit{Spin-orbit Coupling Effects in Two-Dimensional Electron and Hole Systems}, Springer (2003).

\bibitem{note1}
At the boundary between the WL and AM regimes, we assign to $E$ the value reported in Ref(\onlinecite{Santander2011}), as we surmize that the field $E$ on the STO side of the interface 
is caused by all the charge carriers whether they are mobile 
or not.

\bibitem{Ueno2008}
K.~Ueno \textit{et~al.}, Nat. Mat. {\bf7}, 855-858 (2008).

\bibitem{Santander2011}
A.~F.~Santander-Syro \textit{et~al.}, Nature {\bf469}, 189 (2011).

\bibitem{Seo2009}
S.~S.~A.~Seo \textit{et~al.}, Appl. Phys. Lett. {\bf95}, 082107 (2009).

\bibitem{BenShalom2010b}
M.~Ben~Shalom, A.~Ron, A.~Palevski, Y.~Dagan, Phys. Rev. Lett. {\bf105}, 206401 (2010).

\bibitem{Caviglia2010b}
A.~D.~Caviglia \textit{et~al.}, Phys. Rev. Lett. {\bf105}, 236802 (2010).

\bibitem{Copie2009}
O.~Copie \textit{et~al.}, Phys. Rev. Lett. {\bf102}, 216804 (2009).

\bibitem{Pentcheva2009}
R.~Pentcheva, W.~E.~Pickett, Phys. Rev. Lett. {\bf102}, 107602 (2009).

\bibitem{Popovic2008}
Z.~S.~Popovi\ifmmode \acute{c}\else \'{c}\fi{}, S.~Satpathy,R.~M.~Martin, Phys. Rev. Lett. {\bf101}, 256801 (2008).

\bibitem{Son2009}
W.-J.~Son, E.~Cho, B.~Lee, J.~Lee, S.~Han, Phys. Rev. B {\bf79}, 245411 (2009).

\bibitem{Ghosez2011}
P.~Delugas, A.~Filippetti, V.~Fiorentini, D.~I.~Bilc, D.~Fontaine, P.~Ghosez, Phys. Rev. Lett. {\bf106}, 166807 (2011).

\bibitem{Shen2011}
W.~Meevasana \textit{et~al.}, Nat. Mat. {\bf10}, 114 (2011).


\bibitem{Dagan2009}
M.~Ben~Shalom, C.W.~Tai, Y.~Lereah, M.~Sachs, E.~Levy, D.~Rakhmilevitch, A.~Palevski, Y.~Dagan, Phys. Rev. B {\bf80}, 140403(R) (2009).

\bibitem{Castellani2001}
R.~Raimondi, M.~Leadbeater, P.~Schwab, E.~Caroti, C.~Castellani, Phys. Rev. B {\bf64}, 235110 (2001).


\end{thebibliography}

\begin{thebibliography}{10}

\bibitem{Maekawa1981}
S.~Maekawa, H.~Fukuyama, J. Phys. Soc. Jpn. {\bf50}, 2516 (1981).

\bibitem{Dyakonov1972}
M.~I.~D'yakonov, V.~I.~Perel', Sov. Phys. Solid State {\bf13}, 3023 (1972).

\bibitem{Simon2010}
E.~Simon, A.~Szilva, B.~Ujfalussy, B.~Lazarovits,G.~Zarand, L.~Szunyogh, 
Phys. Rev. B {\bf 81}, 235438 (2010).

\bibitem{Winkler2003}
R.~Winkler, \textit{Spin-orbit Coupling Effects in Two-Dimensional Electron and Hole Systems}, Springer (2003).

\bibitem{Ghosez2011}
P.~Delugas, A.~Filippetti, V.~Fiorentini, D.~I.~Bilc, D.~Fontaine, P.~Ghosez, 
Phys. Rev. Lett. {\bf106}, 166807 (2011).

\bibitem{Santander2011}
A.~F.~Santander-Syro \textit{et~al.}, Nature {\bf469}, 189 (2011).

\end{thebibliography}

\end{document}